\documentclass[fleqn,10pt]{wlscirep}
\usepackage{relsize}
\usepackage{graphicx}
\usepackage{amsmath}
\usepackage{hyperref}
\usepackage{tabularx}
\usepackage{adjustbox}

\title{Enhanced spectral profile in the study of Doppler-broadened Rydberg atoms}

\author[1]{Bo-Han Wu}
\author[1]{Ya-Wen Chuang}
\author[1,+]{Yi-Hsin Chen}
\author[1]{Jr-Chiun Yu}
\author[2]{Ming-Shien Chang}
\author[1,*]{Ite A. Yu}
\affil[1]{Department of Physics and Frontier Research Center on Fundamental and Applied Sciences of Matters, National Tsing Hua University, Hsinchu 30013, Taiwan}
\affil[2]{Institute of Atomic and Molecular Sciences, Academia Sinica, Taipei 10617, Taiwan}
\affil[+]{Corresponding author: yhchen920@gmail.com}
\affil[*]{Corresponding author: yu@phys.nthu.edu.tw}

\begin{abstract}
Combination of the electromagnetically-induced-transparency (EIT) effect and Rydberg-state atoms has attracted great attention recently due to its potential application in the photon-photon interaction or qubit operation. In this work, we studied the Rydberg-EIT spectra with room-temperature $^{87}$Rb atoms. Spectroscopic data under various experimental parameters all showed that the contrast of EIT transparency as a function of the probe intensity is initially enhanced, reaches a maximum value and then decays gradually. The contrast of spectral profile at the optimum probe field intensity is enhanced by $2-4$ times as compared with that at weak intensity. Moreover, the signal-to-noise ratio of the spectrum can potentially be improved by 1 or 2 order of magnitude. We provided a theoretical model to explain this behavior and clarified its underlying mechanism. Our work overcomes the obstacle of weak signal in the Rydberg-EIT spectrum caused by an apparent relaxation rate of the Rydberg polariton and weak coupling transition strength, and provides the useful knowledge for the Rydberg-EIT study.
\end{abstract}
\begin{document}

\flushbottom
\maketitle

\thispagestyle{empty}
\newcommand{\FigOne}{
\begin{figure}[t]
\centering
\includegraphics[width=12cm]{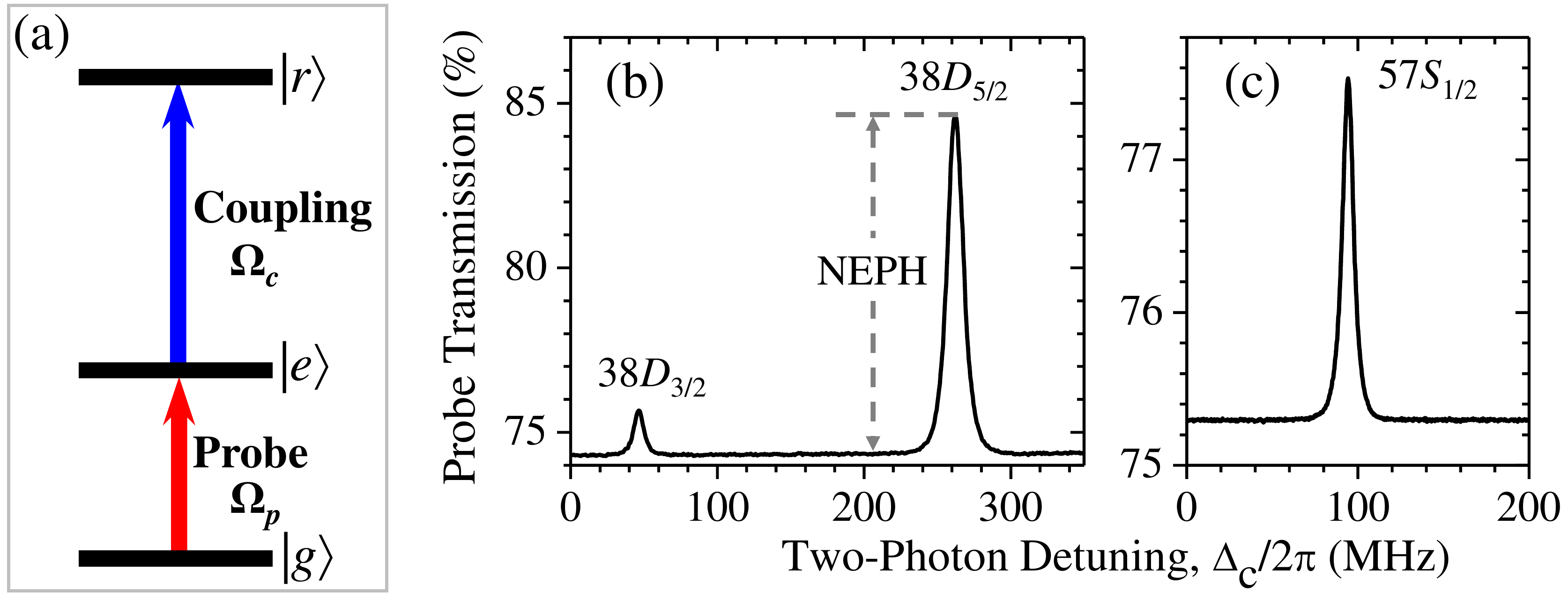}
	\caption{(a) Rydberg-EIT transition scheme. $|g\rangle$ is the ground state $|5S_{1/2}, F=2\rangle$; $|e\rangle$ is the intermediate state $|5P_{3/2}, F'=3\rangle$; and $|r\rangle$ is the Rydberg state $|nS\rangle$ or $|nD\rangle$ according to the experimental measurement. (b) and (c) Rydberg-EIT spectra. The coupling field frequency was swept across the transitions of $|5P_{3/2},F'=3\rangle$ to $|38D_{3/2},F''=2,3\rangle$ and $|38D_{5/2},F''=2,3,4\rangle$ in (b), and across the transition of that to $|57S_{1/2},F=2\rangle$ state in (c), while the probe field frequency was fixed. In (b), we provide the definition of the normalized EIT peak height (NEPH), which is the difference between the transmissions of EIT peak and baseline. The intensities of the probe and coupling fields were 0.029 and 18 W/cm$^2$, respectively. Both light fields had the same polarization $\sigma_{+}$ in (b); and the probe field was $\sigma_{-}$ polarized and coupling field was $\sigma_{+}$ polarized in (c). The values of NEPH are 0.014, 0.10, and 0.022 from left to right peaks in (b) and (c).}
	\label{fig:scheme}
	\end{figure}
}


\newcommand{\FigTwo}{
\begin{figure}[b]
\centering\includegraphics[width=12cm]{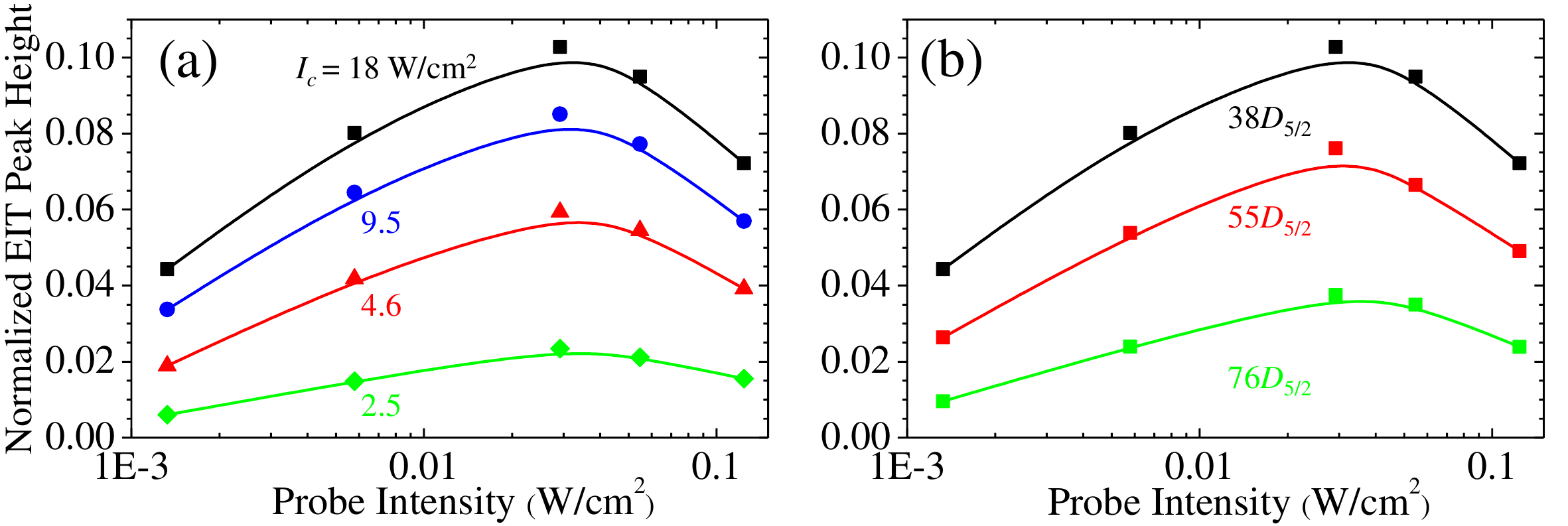}
\caption{NEPH as a function of the intensity of the input probe field at (a) different intensities of the coupling field for the transition to Rydberg state of $|38D_{5/2}\rangle$ and (b) different principal quantum numbers of Rydberg states for the coupling intensity of 18 W/cm$^2$. Both measurements have the same polarization configuration of $\sigma_{+}-\sigma_{+}$. The coupling intensities in units of W/cm$^2$ in (a) and selected Rydberg states in (b) are shown in the legends. Solid lines are the curves to guide the eye.}
	\label{fig:coupling power}
\end{figure}
}

\newcommand{\FigThree}{
\begin{figure}[t]
\centering\includegraphics[width=12cm]{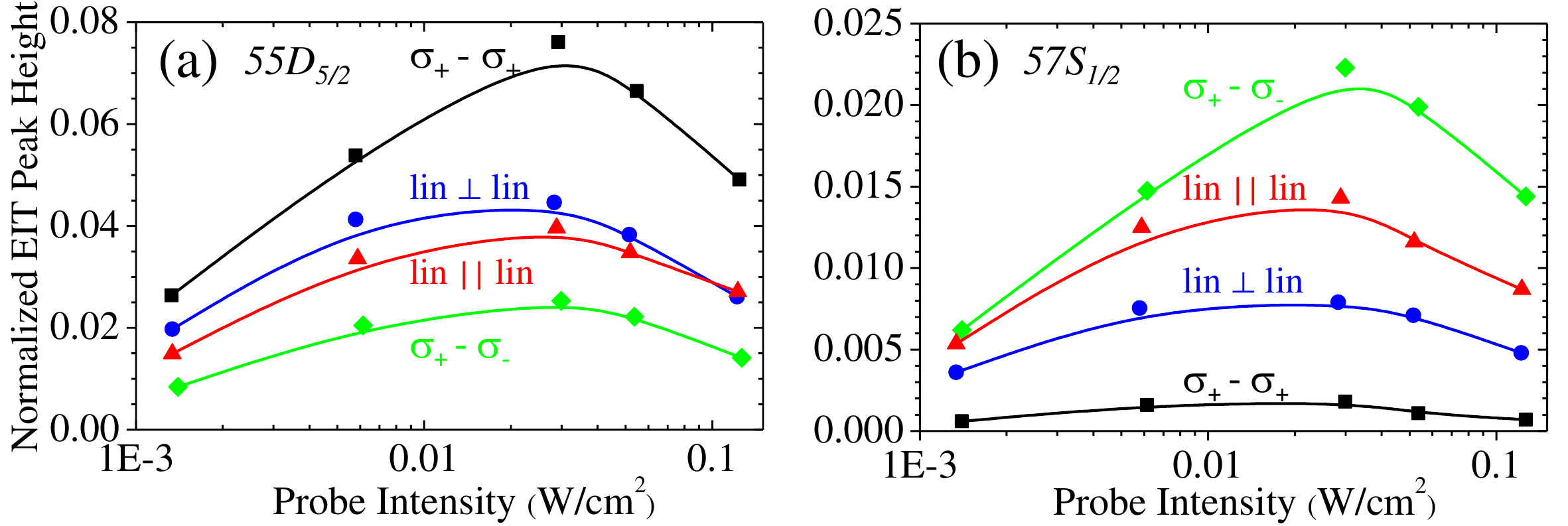}
\caption{NEPH as a function of the intensity of the input probe field at different polarization configurations and Rydberg states, shown in the legends. The coupling intensity was 18 W/cm$^2$ in the both measurements. Solid lines are the curves to guide the eye.}
\label{fig:polarization}
\end{figure}
}
\newcommand{\Table}{
\begin{table}[b]
\centering
\caption{Estimated $\Omega_{c,p}/2\pi$ (in units of MHz) for the coupling intensity of 18 W/cm$^2$ and for the probe intensity of 0.1 W/cm$^2$ in different polarization configurations. We assume the populations are equally distributed among different degenerate Zeeman states of $|g\rangle$. In the final column, $\Omega_{c,p}$ with CGC=1 is present. $\Omega_{p,\rm 0.1W/cm^2}$ is shown in the final row.}
\begin{tabular}{| c | c | c | c | c | c |} 
\hline
{Transitions}
& $\sigma_{+}-\sigma_{+}$ &  lin $\perp$ lin &  lin $\parallel$ lin &  $\sigma_{+}-\sigma_{-}$& $\rm CGC=1$\\ [0.5ex]
\hline
$|5P_{3/2},F'=3\rangle$ to $|38D_{5/2}\rangle$  & 5.1& 3.9 & 4.2& 2.6& 5.7\\ 
\hline
 $|5P_{3/2},F'=3\rangle$ to $|55D_{5/2}\rangle$  & 2.9 & 2.2& 2.4 & 1.5& 3.2\\
\hline
$|5P_{3/2},F'=3\rangle$ to $|57S_{1/2}\rangle$  & 0.20& 0.49& 0.69& 0.82& 1.7\\
\hline
$|5P_{3/2},F'=3\rangle$ to $|76D_{5/2}\rangle$ & 1.8&1.4 &1.5 &0.90& 2.0\\
\hline
\hline
$|5S_{1/2},F=2\rangle$ to $|5P_{3/2},F'=3\rangle$ \hspace{0.2cm} & 22.7&22.7 &22.7 &22.7& 33.2\\
\hline
\end{tabular}
\label{Table}
\end{table}
}

\newcommand{\FigFour}{
\begin{figure}[t]
\centering\includegraphics[width=16cm]{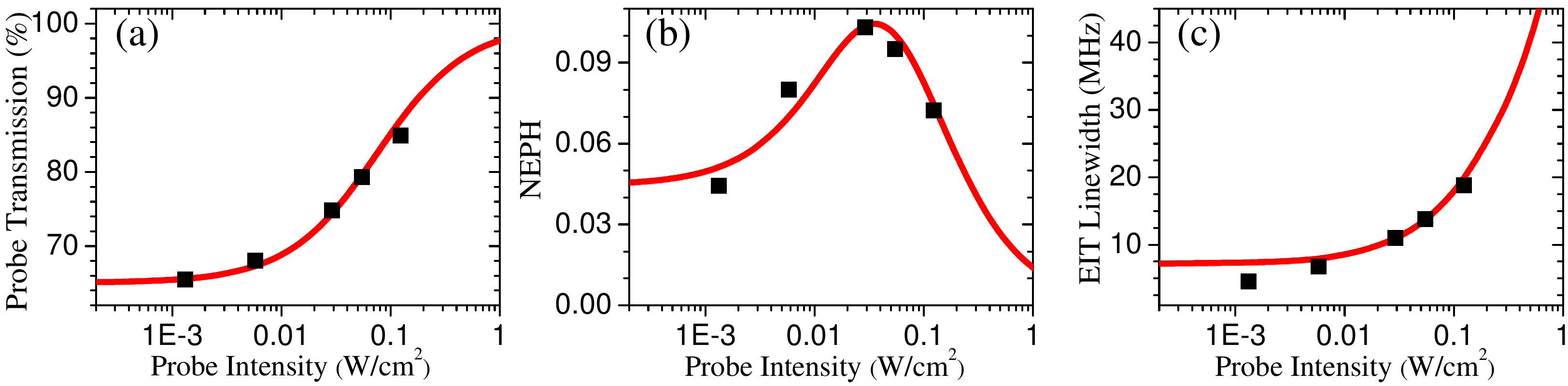}
\caption{(a) Baseline transmission, i.e. the probe transmission at the coupling field frequency detuned far away from the two-photon resonance, (b) NEPH, and (c) linewidth, i.e. full width at half maximum, of the EIT profile versus the intensity of the input probe field. Black squares are the experimental data of the Rydberg state of $|38D_{5/2}\rangle$, and red lines are the theoretical predictions. The coupling intensity was 18 W/cm$^2$, and two light fields were both $\sigma_+$-polarized in the experimental measurement. We set $\Gamma_r = 2\pi\times 5$ kHz in the theoretical calculation. To fit the data well, $\alpha = 0.45$ and $(\Omega_{p,{\rm 0.1W/cm^2}}, \Omega_c, \gamma, \Gamma_e) = 2\pi\times (17.1,~4.5,~3.1,~18)$ MHz. We first determined $\alpha$ and the ratio of $\Omega_{p,{\rm 0.1W/cm^2}}$ to $\Gamma_e$ by the fitting of (a). Then, the ratios of $\Omega_c$ and $\gamma$ to $\Gamma_e$ were resolved by the fitting of (b). Finally, we settled $\Gamma_e$ by the fitting of (c).}
	\label{fig:three fitting}
\end{figure}
}


\newcommand{\FigFive}{
\begin{figure}[t]
\centering
\includegraphics[width=16cm]{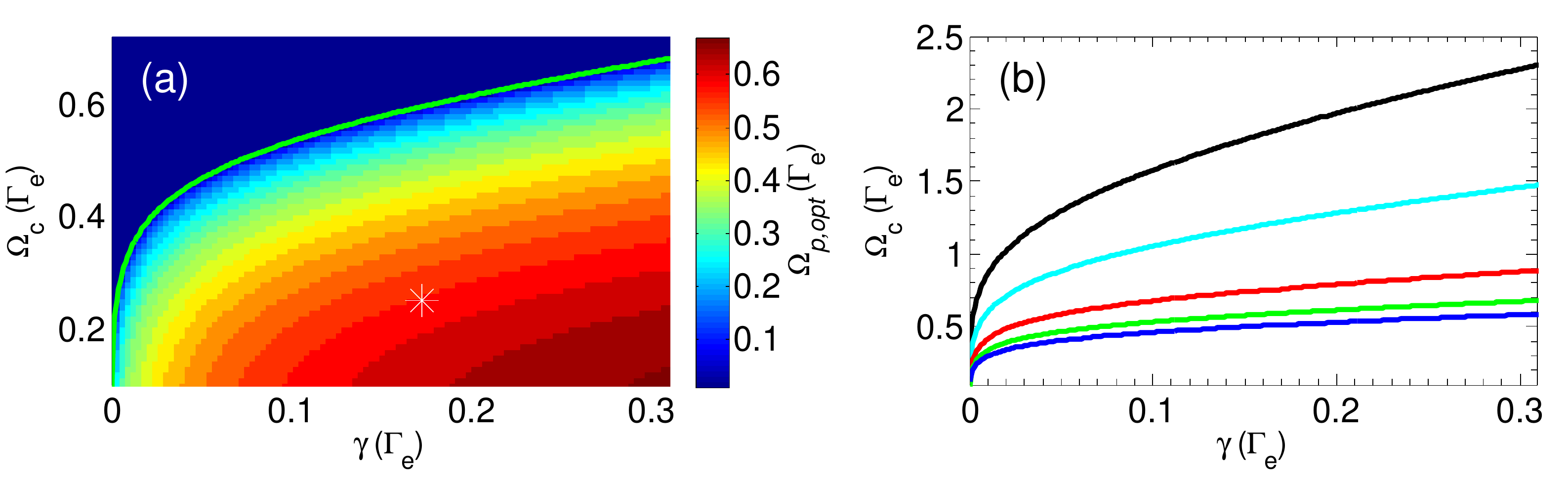}
	\caption{Panel (a) shows the simulation results of $\Omega_{p,\rm opt}$ under a given set of $\Omega_c$ and $\gamma$, where $\Omega_{p,\rm opt}$ is the probe Rabi frequency making NEPH reach the maximum value. OD is set as 0.45 for the calculation. The green solid line represents the boundary between zero and positive values of $\Omega_{p,\rm opt}$. The marker $\ast$ shows the experimental condition of Fig.~\ref{fig:three fitting}. (b) Boundary lines for different optical densities. From top to bottom, ODs are 3 (black), 2 (cyan), 1 (red), 0.45 (green), and 0.1 (blue).}
	\label{fig:contour}
\end{figure}
}

\section*{Introduction}

Rydberg atom has become a popular research topic in recent decades, especially in the context of quantum information science, thanks to its physical properties. The weak dipole transition between ground and highly-excited Rydberg states prolongs the lifetime of Rydberg atoms~\cite{Saffman2010,Gallagher1988}. The large polarizability of Rydberg atoms gives rise to strong long-range interactions. It would couple the nearby atoms strongly through the immense dipole-dipole interaction. The strong interaction between Rydberg atoms leads to a blockade effect, implying a double excitation for a distance smaller than the blockade radius is strongly suppressed~\cite{Comparat2010,Sevincli2011,Friedler2005,Gorshkov2011}. With the unique features of above, Rydberg atom is a good candidate for the demonstration of novel quantum devices, such as single-photon transistors~\cite{Tiarks2014,Gorniaczyk2014} as well as quantum phase gate~\cite{Isenhower2010,Shahmoon2011,Tiarks2016}, single-photon sources~\cite{Saffman2002,Dudin2012,Muller2013}, and quantum simulator~\cite{Weimer2010}.  

Electromagnetically-induced-transparency (EIT) spectrum provides the direct nondissipative optical detections of Rydberg energy levels, atom-atom interaction, and wall-atom interaction in a thin cell~\cite{Mohapatra2007,Pritchard2010,Kubler2010,Pritchard2013,Low2012}. An additional microwave field can break the symmetry of Rydberg-EIT interference, making it a good way to precisely determine the electric field of the microwave~\cite{Sedlacek2012}. Besides the EIT spectra, the quantum information carried by photons can be dynamically encoded in Rydberg polaritons, allowing for storage, control, and retrieval of quantum states~\cite{Maxwell2013,Ripka2016}. To perform the above mentioned studies with Rydberg-state atoms, it is necessary to lock laser frequencies to a two-photon transition frequency. The EIT spectrum provides a convenient way to stabilize the laser frequencies for the studies~\cite{Abel2009}. 

We carried out our study in a room-temperature $^{87}$Rb atomic vapor and investigated the Rydberg EIT spectra in this paper. The EIT peak height, i.e. the difference between the probe transmission at the EIT peak and that in the absence of the EIT effect, has been enhanced by $2-4$ times at the optimum probe intensity as compared with that at weak probe intensity. Remarkably, the optimum intensity is influenced very little by the light polarization, orbitals of Rydberg state, and the principal quantum number of Rydberg state, $n$. 
We will provide a theoretical model for the observed behavior of the EIT peak height as a function of the probe intensity. 
In addition, when one applies a stronger probe intensity or power in the measurement, the signal level of the probe field is immediately enhanced. Thus, as for dominant noise being not caused by fluctuation of the probe power or intensity (but being caused by, for examples, stray light, electronic noise, detector's dark current, etc.), the signal-to-noise ratio ($S/N$) can be significantly improved. On the other hand, the spectral linewidth increases only by 2 folds. Therefore, the comprehensive feature of EIT effect leads to a better way for locking the upper transition frequency through a high contrast of the Rydberg-EIT spectrum, making it useful for the Rydberg-relevant researches.

\FigOne
\section*{Results and Discussion}
\subsection*{Experimental Observation}
Rydberg EIT has a cascade-type ($\Xi$-type) level structure, which consists of a ground state, an intermediate excited state, and a Rydberg state, as shown in Fig.~\ref{fig:scheme}(a). 
The probe field couples the ground state $|5S_{1/2}, F=2\rangle \equiv |g\rangle $ and the intermediate state $|5P_{3/2},F'=3\rangle \equiv |e\rangle$, while the coupling field drives the transition of $|e\rangle$ to Rydberg state $|r\rangle$. For each EIT spectrum measurement, we kept the probe field frequency resonant to $|g\rangle$-$|e\rangle$ transition, while the frequency of the coupling field was swept across the $|e\rangle$-$|r\rangle$ transition.
The probe field has the maximum transmission when the two light fields are resonant to the Rydberg state due to the quantum interference of EIT effect. Figures~\ref{fig:scheme}(b) and~\ref{fig:scheme}(c) show the typical Rydberg-EIT spectra. Far from resonance, atoms in the excited state are less likely pumped to the Rydberg state. It can be viewed as a two-level transition, resulting a high absorption. Here, we define the normalized EIT peak height (NEPH) as the transmission difference between EIT and two-level transitions. The values of NEPH are 0.014, 0.10, and 0.022 from left to right peaks in Figs.~\ref{fig:scheme}(b) and~\ref{fig:scheme}(c). \FigTwo

To systematically study the NEPH, we varied the intensities of the probe and coupling fields, the Rydberg states, and the light polarization configurations for different orbitals of Rydberg state. As shown in Fig.~\ref{fig:coupling power}, a stronger $\Omega_c$ (with fixed Rydberg state of $|38D_{5/2}\rangle$) or a lower Rydberg state (with fixed coupling intensity of 18~W/cm$^2$) leads to a greater NEPH. Note that the intensity of the coupling or probe field specified in this article is the intensity at the center of the Gaussian beam profile of the laser field. The phenomenon can be explained by the expression of EIT transmission $ T\sim$ $\rm Exp[{-2\alpha \gamma\Gamma/\Omega_c^2}]$ under perturbation limit of the probe field, i.e. the Rabi frequency of probe field $\Omega_p$ is much weaker than that of the coupling field $\Omega_c$~\cite{Fleischhauer2005,YFC2006}. Here $\alpha$ is the optical density of the medium, $\gamma$ is the decoherence rate between Rydberg and ground states, and $\Gamma$ is the spontaneous decay rate of the intermediate state. The coupling Rabi frequency $\Omega_c$ is proportional to a power law $n^{*-3/2}$, where $n^*$ is the effective principal quantum number of Rydberg state (see Method). The coupling strength between a higher Rydberg state and the intermediate state is weaker, resulting in a smaller $\Omega_c$. The expression shows that a stronger $\Omega_c$ leads to a higher transmission of the probe field, which can qualitatively describe the data. 

Besides, the polarization configurations of the laser fields were adjusted as $\sigma_{+} - \sigma_{+}$, lin $\perp$ lin, lin $\parallel$ lin, and $\sigma_{+} - \sigma_{-}$ by the half- or quarter-wave plates. For $|nD_{5/2}\rangle$ Rydberg state, the best polarization configuration is $\sigma_{+} - \sigma_{+}$ (Fig.~\ref{fig:polarization}(a)). The hyperfine levels $F'' = 2,3,4$ of $|nD_{5/2}\rangle$ state are not resolvable so that all transition channels among the Hyperfine states and their Zeeman sub-levels need to be taken into account. The Zeeman state transition $|F'=3, m_F' = 3\rangle$ to $|F''=4, m_F'' = 4\rangle$ is a cycling transition, leading to a larger effective $\Omega_c$. On the contrary, when the coupling field couples to $|nS_{1/2}\rangle$ Rydberg state, the polarization of $\sigma_{+} - \sigma_{-}$ is the best configuration shown in Fig.~\ref{fig:polarization}(b). For $|nS_{1/2}\rangle$ state, only the excitation to $F''=2$ hyperfine state is allowed. 
For example, in the polarization configuration of $\sigma_{+} - \sigma_{+}$ there are some probe excitations do not couple with the coupling field and, hence, the EIT transmission is reduced. Therefore, in order to get the best EIT contrast for a given $|nS\rangle$ or $|nD\rangle$ state, one should well select the best light polarization.  
\FigThree

For each light polarization configuration or each Rydberg state, $n$, as well as the orbital, $S$ or $D$, of Rydberg state, the Rabi frequencies of light shall be modified by averaging Clebsch-Gordan coefficients (CGC) among different hyperfine and Zeeman states.
We define $\Omega_{p,\rm 0.1W/cm^2}$ as the Rabi frequency of the probe field with intensity of 0.1W/cm$^2$ and $\Omega_{p,\rm 0.1W/cm^2} = 2 \pi \times 22.7$ MHz for each polarization of $\sigma_+$, or $\sigma_-$, or $\pi$. The Rabi frequencies used in the measurements are summarized in Table~\ref{Table}. A detailed description of the derivation of Rabi frequencies can be found in Method. From the measurements of Figs.~\ref{fig:coupling power} and ~\ref{fig:polarization} and the analysis of Rabi frequencies, a greater NEPH corresponding to a larger $\Omega_c$ is quantitative proved. 
\Table

Moreover, all measured results reveal a universal phenomenon, that as we enlarged the probe field intensity, NEPH initially increased, then reached a maximum (around intensity of $0.03~\rm W/cm^2$), and finally decreased. The optimum probe power is about 20 times stronger than the weakest power used in the measurement so that the $S/N$ of NEPH can be enhanced. 
In most of EIT-relevant studies, people typically applied a weak probe intensity or Rabi frequency ($\Omega_p \ll \Omega_c$) because EIT theories work well under the perturbation limit. To our knowledge, this rising phenomenon of NEPH is not presented in any article so far and promotes a further study in this paper.
After the optimization of the probe intensity, the best NEPH is $2-4$ times larger, and therefore, the overall $S/N$ can have $1-2$ order of magnitude enhancement. Qualitatively, such enhancement or the behavior of rising NEPH is universal, which is irrelative to the laser polarization, the coupling intensity, the orbital and principal quantum number $n$ of Rydberg state. The physics mechanism and criterion or experimental condition of the behavior will be discussed in the remaining part of the Results and Discussion. 
\FigFour


\subsection*{Theoretical Model} 
We present the theoretical simulation by solving the optical Bloch equations (OBEs) and Maxwell-Schrödinger equation (MSE). According to the EIT theoretical studies~\cite{Su2011,Su2011PRA}, the spectrum or the dynamic behaviors in Doppler-broadened media can be simulated by a simple model used for Doppler-free media with the modifications of parameters, such as decoherence rate $\gamma$, optical depth $\alpha$, and Rabi frequencies $\Omega_{c,p}$. 
Hence, the following equations are widely used in the EIT-relevant studies,
\begin{subequations}
\begin{align}
\partial_{t}\rho_{eg}&=\frac{i}{2}\left[\Omega_{p}\left(\rho_{gg}-\rho_{ee}\right)+\Omega_{c}\rho_{rg}\right]-\frac{\Gamma_{e}}{2}\rho_{eg}, \\  
\partial_{t}\rho_{er}&=\frac{i}{2}\left[\Omega_{p}\rho_{rg}^*+\Omega_{c}\left(\rho_{rr}-\rho_{ee}\right)\right]-\left(\frac{\Gamma_{e}+\Gamma_{r}}{2}+i\Delta_{c}\right)\rho_{er}, \\
\partial_{t}\rho_{rg}&=\frac{i}{2}\left(\Omega_{c}^*\rho_{eg}-\Omega_{p}\rho_{er}^{*}\right)-\left(\gamma-i\Delta_{c}\right) \rho_{rg}, \\ 
\partial_{t}\rho_{gg}&=\frac{i}{2}\left(\Omega_{p}^*\rho_{eg}-\Omega_{p}\rho_{eg}^{*}\right)+\Gamma_{e}\rho_{ee}, \\  
\partial_{t}\rho_{rr}&=\frac{i}{2}\left(\Omega_{c}^*\rho_{er}-\Omega_{c}\rho_{er}^{*}\right)-\Gamma_{r}\rho_{rr},\\  
1&=\rho_{gg}+\rho_{ee}+\rho_{rr},\\  
&\left(\frac{1}{c}\partial_{t}+\partial_{z}\right)\Omega_{p}=i\frac{\alpha \Gamma_{e}}{2L}\rho_{eg}.  
\end{align}
	\label{eq2}
\end{subequations}
Here $\rho_{ij}$ is an element of the density-matrix operator of the three-level system, $\Delta_{c}$ is the detuning of the coupling field, $\Gamma_{e}$ and $\Gamma_{r}$ are the linewidths or the spontaneous decay rates of the intermediate state and Rydberg state, and $L$ is the length of the medium.  

We will describe how the predictions from above theoretical model can be consistent with the data. We first determine the optical density of the medium $\alpha$ and the ratio of $\Omega_{p,\rm 0.1W/cm^2}$ to $\Gamma_e$ by numerically fitting the data of the probe transmission at the coupling frequency detuned far away from the two-photon resonance versus the intensity of the input probe, as shown in Fig.~\ref{fig:three fitting}(a). The transmission at small values of $\Omega_p$ determines the value of optical density mainly. As $\Omega_p$ gets larger, the power broadening effect, indicated by $\Omega_p^2/\Gamma_e^2$, influences the increment of the transmission. Hence, the best fit gives the optical depth $\alpha=0.45$ and $\Omega_{p,{\rm 0.1W/cm^2}}/\Gamma_e = 0.95$.
Next, we fix $\Gamma_r = 2\pi\times 5$ kHz, which is the spontaneous decay rate of $38D_{5/2}$ \cite{Branden2010}, in the calculation of the NEPH predictions (Fig.~\ref{fig:three fitting}(b)). The ratios of $\Omega_c$ and $\gamma$ to $\Gamma_e$ were resolved by the fitting because both of $\Omega_c$ and $\gamma$ can affect the EIT peak height. This can be realized from the EIT peak transmission $T\sim$ $\rm Exp[{-2\alpha (\gamma/\Gamma_e)/(\Omega_c/\Gamma_e)^2}]$ under perturbation limit of the probe field~\cite{Fleischhauer2005,YFC2006}. In addition, as all the frequency-related quantities, $\Omega_p$, $\Omega_c$, $\gamma$, etc., are normalized to or divided by $\Gamma_e$ in Eq.~(\ref{eq2}), the calculation result becomes invariant with respect to $\Gamma_e$. Therefore, the NEPH can determine $\Omega_c/\Gamma_e$ (or $\gamma/\Gamma_e$) but not $\Omega_c$ and $\Gamma_e$ (or $\gamma$ and $\Gamma_e$) individually. 
The actual value of $\Gamma_e$ in units of MHz was derived by matching the theoretical EIT linewdiths to the experimental ones as illustrated in Fig.~\ref{fig:three fitting}(c). The three kinds of data shown in Figs.~\ref{fig:three fitting}(a), ~\ref{fig:three fitting}(b), and ~\ref{fig:three fitting}(c) can uniquely determine a set of $\alpha = 0.45$ and  $(\Omega_{p,{\rm 0.1W/cm^2}}, \Omega_c, \gamma, \Gamma_e) = 2\pi\times (17.1,~4.5,~3.1,~18)$ MHz. 

The best fit of EIT spectra gives $\Gamma_e = 2\pi\times$ 18 MHz, indicating that only a small fraction of room-temperature atoms interacts with the probe field. Note that the actual angular natural linewidth of the intermediate state $|e\rangle$ is $\Gamma = 2\pi\times$ 6 MHz. In each measurement of the EIT spectrum, we kept the probe field frequency fixed at the resonant frequency of the transition from $|5S_{1/2},F=2\rangle$ to $|5P_{3/2},F'=3\rangle$, and swept the coupling field frequency. Atoms with high velocities $v$ do not interact with the probe field at all due to the Doppler shift $k v$ being large, where $k$ is the wave vector of the probe light. Because the EIT scheme is formed by the cascade- or ladder-type transitions, no interaction with the probe field also means no influence from the coupling field. Only the atoms with low velocities can interact with the probe field and, consequently, also be influenced by the coupling field. In the lab frame, these atoms with low but non-zero velocities possess different resonance frequencies of the transition. A distribution of resonance frequencies, arising from the atoms with various velocities distributed around $v = 0$, can equivalently be seen as the broadening of the excited-state linewidth in the theoretical model treating all atoms with $v = 0$. Therefore, $\Gamma_e$ of $2\pi\times$ 18 MHz indicates that only the atoms, with $|k v|$ less than approximately $2\pi\times$ 9 MHz, interacted with the probe and coupling fields and participated in the measurements.

The EIT linewidth is predominately determined by the coupling field intensity and optical density of the medium at the weak probe field regime. Employing room-temperature or hot atomic media, one typically applies a strong coupling field to diminish the influence of non-negligible decoherence rate $\gamma$ on the EIT transmission. For a strong probe field, the power broadening effect occurs, resulting in the spectral width becomes broader. The linewidth at the optimum probe intensity, shown in Fig.~\ref{fig:three fitting}(c), is only twice broader than that at weak intensity. As discussed before, the overall $S/N$ of EIT peak can have $1-2$ order of magnitude enhancement. The EIT dispersion, i.e. the slope of a frequency-modulated EIT spectrum, becomes much larger. If the probe field is further stronger, the increment of EIT linewidth and the decrement of NEPH becomes fast. The EIT spectrum doesn't benefit the Rydberg studies relied on the optical frequency lock.   

\subsection*{NEPH Behavior}
We now discuss the universal phenomena, that NEPH increases initially with weak $\Omega_p$ regime and then decreases at strong $\Omega_p$ regime. The theoretical model for Doppler-free atoms with effective parameters can phenomenologically describe the Doppler-broadened atomic medium well. Hence, in the following discussion, we will utilize this model to explain the observed behavior of NEPH. 
For simplicity, we start from the analysis of the absorption cross sections, which is equal to $\sigma_0 \times {\rm Im}[\rho_{eg} \Gamma_e / \Omega_p]$,  where $\sigma_{0}$ is the resonant absorption cross section under perturbation limit of the probe field. We define $\sigma_{TL}$ as the absorption cross section of the baseline in a spectrum, which is resulted from the resonant transition in the two-level system. $\sigma_{EIT}$ is defined as the absorption cross section of the EIT peak.
Based on the steady-state solution of the optical-Bloch equations, we derive $\sigma_{TL}$ and $\sigma_{EIT}$ under the assumptions of $\Gamma_r \ll \Gamma_e$ and $\Gamma_r\Gamma_e  \ll \Omega_c^2$ as
\begin{equation}
\frac{\sigma_{TL}}{\sigma_{0}}  = \frac{\Gamma_e^2}{2\Omega_{p}^2+\Gamma_{e}^2},\\
\frac{\sigma_{EIT}}{\sigma_{0}}  = \frac {\Gamma_e^2\left (2\gamma\Omega_c^2 + \Gamma_r\Omega_p^2 \right)}{\Gamma_e\Omega_c^2\left (\Omega_c^2 + 2 \gamma \Gamma_e \right) + \Omega_p^2\left(6\gamma\Omega_c^2 + \Gamma_e\Omega_c^2 + 2\Gamma_r\Omega_p^2 \right)}.
\label{eq:sigma}
\end{equation}

The analysis of absorption cross section as well as the probe transmission will be discussed in two regimes: weak and strong $\Omega_{p}$ regimes.
In the weak $\Omega_{p}$ regime, we further assume $\Omega_p^2 \ll \Gamma_{e}^2, \Omega_c^2$ and $\Gamma_r \Omega_p^2 \ll \Gamma_e \Omega_c^2$. The absorption cross sections of two-level and Rydberg-EIT systems can be expanded as
\begin{equation}
\frac{\sigma_{TL}}{\sigma_{0}}  \approx 1-\frac{2 \Omega_{p}^2}{\Gamma_{e}^2},\\
\frac {\sigma_{EIT}} {\sigma_0} \approx \frac{2\gamma\Gamma_e} 
{2\gamma\Gamma_e+\Omega_c^2}-\frac{2 \Omega_p^2\gamma \left(6\gamma+\Gamma_e\right)}{\left(\Omega_c^2+2\gamma\Gamma_e\right)^2}, \\
\Delta\sigma  = \sigma_{TL}-\sigma_{EIT} = A_1+\frac{2B_1\Omega_p^2}{\Gamma_e^2},\\
\label{smallOmegaP}
\end{equation}
where 
\begin{equation}
\nonumber
B_1= \frac{\gamma  \Gamma_e^2 (2 \gamma +\Gamma_e)-\Omega_c^2 \left(4 \gamma \Gamma_e+\Omega_c^2\right)}{\left(\Omega_c^2 + 2 \gamma \Gamma_e\right)^2}.
\end{equation}
If $B_1$ is positive, $\Delta\sigma$ as well as NEPH increases with increasing $\Omega_p$. In the physical explanation, on one hand, the transmission of the probe field under the two-level transition slightly increases as increasing $\Omega_p$ due to the power broadening effect. On the other hand, in the EIT transition, a highly-excited Rydberg state can be treated as a meta-stable state because of the slow spontaneous decay rate. A strong $\Omega_p$ results in more populations to be driven into the excited state $|e\rangle$ and Rydberg state $|r\rangle$. According to EIT theory, the population ratio of $|r\rangle$ to $|g\rangle$ is determined by the ratio of $\Omega_p^2$ to $\Omega_c^2$, and, thus, the population ratio becomes larger as $\Omega_p^2$ increases under fixed $\Omega_c^2$. The less population in the ground state leads to a higher transmittance of the probe field. With a weak $\Omega_c$, the rising transmittance for EIT transition would be faster than that for two-level transition. The initially-rising phenomenon in the data of NEPH versus the probe intensity can occur under $B_1 > 0$

We now intend to explain the decreasing behavior at stronger $\Omega_p$ regime. Here, we assume $\Omega_p^2 \gg \Gamma_e^2, \Omega_c^2$.
$\sigma_{TL}$ and $\sigma_{EIT}$ are presented as
\begin{equation}
	\label{eq4}
\frac{\sigma_{TL}}{\sigma_{0}}  \approx \frac{\Gamma_{e}^2}{2 \Omega_{p}^2},\\
\frac{\sigma_{EIT}}{\sigma_0}  \approx \frac{\Gamma_e^2\left(2\gamma\Omega_c^2+\Gamma_r\Omega_p^2\right)}{\Omega_p^2\left[\left(6\gamma+\Gamma_e\right)\Omega_c^2+2\Gamma_r\Omega_p^2\right]},\\
\Delta\sigma  = \sigma_{TL}-\sigma_{EIT}  = \frac {A_2\Gamma_e^2}{2\Omega_p^2},
\end{equation}
where
\begin{equation}
\nonumber
A_2 = \frac{\left(2\gamma + \Gamma_e\right)\Omega_c^2}{\left(6 \gamma+\Gamma_e\right)\Omega_c^2+2\Gamma_r \Omega_p^2}.
\end{equation}
$A_2$ is always larger than zero, implying that $\Delta\sigma$ as well as NEPH decreases with increasing $\Omega_p$. 
At a large $\Omega_p$, the EIT peak transmission becomes saturated and has a little room for improvement, while the baseline transmission can still be significantly increased. Consequently, increment of the probe Rabi frequency makes the baseline transmission approach to the EIT peak transmission, implying the NEPH decreases to zero. 
The above argument is supported by $\Delta \sigma$ decreasing monotonically with $\Omega_p$ as shown by Eq.~(\ref{eq4}). Since NEPH initially increases at small $\Omega_p$ and finally decreases at large $\Omega_p$, there must exist an optimum probe Rabi frequency, $\Omega_{p,{\rm opt}}$, which makes NEPH reach its maximum value.  

\FigFive

To observe the phenomenon that NEPH as a function of the probe intensity initially increases, then reaches a maximum, and eventually decreases to zero, $B_1$ must be positive. The condition $\Omega_{c}^2 \le \Gamma_e\left(\sqrt[]{6\gamma^2+\Gamma_e\gamma}-2\gamma\right)$ makes $B_1 > 0$. In the numerical calculation with a given set of $\Omega_c$ and $\gamma$, we search for the optimal probe Rabi frequency, $\Omega_{p,\rm opt}$, as shown in Fig.~\ref{fig:contour}(a). The solid line sets the boundary in the parameter space between regions with zero and positive $\Omega_{p,\rm opt}$. 
The experimental condition of the measurements in Fig.~\ref{fig:three fitting} were far from the border line, shown as $\ast$ in the contour plot. Note that for a sufficient small optical density, e.g. OD $<$ 0.1, the border line (blue line in Fig.~\ref{fig:contour}(b)) well fits the above-mentioned condition. As the OD gets larger, the increasing NEPH phenomenon is within reach for a sufficient weak $\Omega_c$, as shown in Fig.~\ref{fig:contour}(b). 
Generally, the enhancement of EIT contrast or the increasing phenomenon of NEPH can be observed in the $\Lambda$-EIT or Rydberg-EIT systems with Doppler-broadened atoms or with cold atoms under an apparent decoherence rate~\cite{Su2011,Su2011PRA}.

\section*{Conclusion}
We systematically investigated the best contrast of EIT peak in different polarization configurations of light fields, laser intensities, and orbitals and principal quantum numbers of Rydberg states. From the measurements and the analysis of Rabi frequencies, a greater NEPH corresponding to a larger $\Omega_c$ is quantitative proved. 
In addition, a universal phenomenon was experimentally observed that the EIT contrast increasing with the probe intensity and then decreasing for further stronger light field.
The EIT contrast is enhanced by $2-4$ times at the optimum probe field intensity as compared with that at weak intensity. Meanwhile, the signal-to-noise ratio of the profile can be improved by 1 or 2 order of magnitude. The enhancement can be achieved in most of the $\Lambda$-EIT or Rydberg-EIT experiments using Doppler-broadened atoms or using cold atoms under an apparent decoherence rate.
Without considering the Doppler effect due to the atomic motion, the thermal Rydberg-EIT spectrum can be predicted by using the simulation for a Doppler-free medium with the modification of parameters. The theoretical model can well fit the data and simulate the rising behavior of NEPH.  
Our studies provide a better way of locking the upper transition frequency through a high contrast Rydberg-EIT spectrum and advances the Rydberg-atom-relevant studies.  

\section*{Methods}
\subsection*{Setup and Measurements}
The probe field couples $|g\rangle$ and $|e\rangle$ states with wavelength of $780$ nm, generated from an external cavity diode laser. The coupling field drives atoms from $|e\rangle$ to $|r\rangle$ states with wavelength of $480$~nm, produced from a frequency doubled diode laser system (Toptica TA SHG pro). These two beams are sent into a counter-propagation direction to diminish the Doppler effect due to the atomic motion. We applied a dichroic mirror (DM) to separate the mixed probe field from the coupling beam after the vapor cell. The full width at $e^{-2}$ maximum of the probe and coupling beams were 1.4 mm and 2.2 mm, respectively. The vapor cell is filled with the admixture of $^{87}$Rb and $^{85}$Rb atoms at the temperature of $300$ K.

\subsection*{Deviation of Table~\ref{Table}}
Considering the equivalent CGC \cite{Guan2007,Deiglmayr2006}, the Rabi frequency is derived as a function of laser intensity, 
\begin{equation}
\Omega_p= \langle a_p\rangle \Gamma~\sqrt[]{\frac{I_{p}}{2 I_{sat}}}, \\
\Omega_c= \frac{e a_0}{\hbar}C_{S,D}\langle a_c\rangle~\sqrt[]{\frac{2 I_{c}}{c\epsilon_0}}n^{*-\frac{3}{2}},
	\label{eq1}
\end{equation}
where $\langle a_p\rangle=\sqrt[]{\sum_{i}P_{i}C^2_{p,i}}$ and $\langle a_c\rangle=\sqrt[]{\sum_{i}P_{i}C^2_{p,i} C^2_{c,i}/\langle a_p \rangle^2}$. $I_{p}$ and $I_{c}$ are the laser intensities of probe and coupling fields, $I_{sat}$ is the saturation intensity of the probe field transition. $n^*$ is the effective principal quantum number of Rydberg state which replaces the true $n$ with the relation of $n^*=n-\delta(n,l,j)$. The quantum defects $\delta(n,l,j)$ of $^{87}$Rb atoms were measured by Mack et al.,~\cite{Mack2011} and for $n>20$ that are around 1.34(1) and 3.13 for $|nD_{3/2, 5/2}\rangle$ and $|nS_{1/2}\rangle$ states, respectively. $C_{S}=4.5$ and $C_{D}=8.5$ for $S$ and $D$ orbitals of Rydberg state. 
In the expressions of $\langle a_p\rangle$ and $\langle a_c\rangle$, $P_{i}$ represents the population in the $i$-th Zeeman ground state, $C_{p,i}$ and $C_{c,i}$ are the CGCs of the probe and coupling transitions for each subsystem. 

The polarization configurations of the laser fields were adjusted as $\sigma_{+} - \sigma_{+}$, lin $\perp$ lin, lin $\parallel$ lin, and $\sigma_{+} - \sigma_{-}$ by the half- or quarter-wave plates. To derive the effective $\Omega_{p}$, we consider all  the transition channels among Zeeman states from $|5S_{1/2}, F=2\rangle$ to $|5P_{3/2}, F'=3\rangle$. The transition of $\sigma_+$, or $\sigma_-$, or $\pi$ polarizations of the probe field all gives $\langle a_p\rangle=\sqrt[]{7/15}$. The coupling field drives atoms from $|5P_{3/2}, F'=3\rangle$ to $|nD_{5/2}, F'' = 2,3,4\rangle$ or $|nS_{1/2}, F'' = 2\rangle$. Note that the $D$ state hyperfine levels are not resolvable, and hence all transition channels need to be taken into account. The values of $\Omega_{c,p}$ are summarized in Table~\ref{Table}.

\section*{Acknowledgements}
This work received the support of the Ministry of Science and Technology of Taiwan under Grant Nos. 105-2119-M-007-004, 105-2923-M-007-002-MY3, and 105-2112-M-007-035-MY2. MSC acknowledges the Academia Sinica for equipment grant. BHW, MSC, and IAY acknowledge many fruitful discussions on this work under the platform sponsored by the Experimental Collaboration Program of National Center for Theoretical Science. All authors thank Dr. Bongjune Kim for comments, and Dr. Artūrs Ciniņš and Dr. Teodora Kirova for discussion.

\section*{Author Contributions Statement}
I.A.Y. conceived the study and designed the experiment. Y.-W.C. measured the spectra supervised by M.-S.C. B.-H.W., Y.-H.C., and J.-C.Y. studied the theory and analyzed the data supervised by I.A.Y. B.-H.W. and Y.-H.C. made the figures and table. The manuscript was written by Y.-H.C. with the help from B.-H.W., I.A.Y., and M.-S.C. 

\section*{Additional Information}
The authors declare no competing financial interests.

\bigskip

\bibliography{RydbergReference}
\end{document}